\documentclass[conference]{IEEEtran}
\IEEEoverridecommandlockouts
\usepackage{amsmath,amssymb,amsfonts}
\usepackage{graphicx}
\usepackage{color}
\usepackage{comment}
\usepackage{cite}
\usepackage{url}
\usepackage{tikz}
\usetikzlibrary{shapes}
\usepackage{pgfplots}
\usepackage{multirow, multicol}
\pgfplotsset{compat=1.18} 
\newcommand{\new}[1]{\textcolor{black}{#1}}

\begin{document}

\title{Modeling, Analysis, and Optimization of Cascaded Power Amplifiers}

\author{\IEEEauthorblockN{Oksana Moryakova}
    \IEEEauthorblockA{\textit{Department of Electrical Engineering} \\
        \textit{Link\"{o}ping University}\\
        Link\"{o}ping, Sweden \\
        oksana.moryakova@liu.se}
    \and
    \IEEEauthorblockN{Thomas Eriksson}
    \IEEEauthorblockA{\textit{Department of Electrical Engineering} \\
        \textit{Chalmers University of Technology}\\
        Gothenburg, Sweden \\
        thomase@chalmers.se}
    \and
    \IEEEauthorblockN{H\aa kan Johansson}
    \IEEEauthorblockA{\textit{Department of Electrical Engineering} \\
        \textit{Link\"{o}ping University}\\
        Link\"{o}ping, Sweden \\
        hakan.johansson@liu.se}
}

\maketitle

\begin{abstract}
This paper deals with modeling, analysis, and optimization of power amplifiers (PAs) placed in a cascaded structure, particularly the effect of cascaded nonlinearities is studied by showing potential ways to minimize the total nonlinearities. The nonlinear least-squares algorithm is proposed to optimize the PA parameters along with the input power level, and thereby minimize the total nonlinearities in the cascaded structure. The simulation results demonstrate that the performance of the optimized configurations for up to five PAs using the proposed framework can improve the linearity properties of the overall cascade.
\end{abstract}


\section{Introduction}
\label{sec:introduction}
Radio frequency power amplifiers (PAs) are essential elements in wireless communication systems, but they are also one of the main sources of nonlinearities that distort the transmitted signal and degrade the overall system performance. This is especially severe in systems operating on signals with high peak-to-average power ratio (PAPR). In order to design high-performance wireless transmitters, the behavioral modelling of PAs has received considerable attention over the last few decades, as it is an essential step in understanding PAs limitations and developing efficient digital signal processing algorithms for optimizing their performance \cite{Hammi_2008, Khawam_2020, Moghaddam_2022, He_2024}.  

Recent advances in wireless communication systems have shown the need for considering a cascade of PAs, where each of them amplifies the incoming signal to compensate for the losses incurred in the preceding fiber \cite{Kong_2024}. A similar problem has been considered in optical free-space communications \cite{Yoon_2005} and optical communications \cite{Wannenmacher_1988} to overcome the impact of the attenuation for long distances.
Additionally, cascaded structures of PAs have been proposed to improve power efficiency and minimize chip size \cite{Hammi_2006, Landin_2014}. However, the effect of cascaded nonlinearities has not been fully studied in these areas, moreover, many papers assumed that the nonlinear effect is negligible.

Further, requirements on implementation complexity and power consumption are growing with the evolution of communication systems. It is usually desirable that a PA operates near saturation region in order to achieve maximum power efficiency, which is especially important in ultra-high-frequency communication systems. There has been a large amount of literature on modeling a single PA and various techniques on linearization, including digital pre-distortion at the transmitter \cite{Aggarwal_2019, Yu_2025} and digital post-distortion at the receiver \cite{Ziv_2016}, 
proposed to achieve a balance between linearity, adequate amplification, and implementation complexity.  
However, in the case of cascaded PAs, the distortions from each amplifier accumulate with those from the preceding stages, leading to severe nonlinear behavior. Moreover, since the PA nonlinearity strongly depends on the input power, the behavior of each PA in the cascade is influenced by the output of the preceding PA. This highlights an important research direction for investigating the behavior of cascaded PAs and discovering potential ways to minimize the effect of the cascaded structure.

In this paper, we address this issue by modeling the cascaded PAs, providing a thorough analysis of total nonlinearities, and optimizing the parameters of the cascade to mitigate the severe effect of the cascaded amplifiers. 
The main contributions of the paper are as follows.
\begin{itemize}
    \item An analysis of the nonlinearities in the cascaded PAs structure is presented, showing potential ways of minimizing the effect of cascading.
    
    \item A nonlinear least-squares (NLS) optimization problem is formulated and solved to minimize the total nonlinearities in the cascaded structure by optimizing the PA gain values and input power.
    
    \item Numerical results demonstrating the performance of the optimized configurations using the proposed framework are presented for up to five PAs in cascade.
\end{itemize}

Following the introduction, Section \ref{sec:system} introduces the model of the cascaded PAs. In Section \ref{sec:analysis}, an analysis of nonlinearities in the cascaded structure is given and the ways of minimizing them are highlighted. The optimization problem to minimize the total nonlinearities is presented in Section \ref{sec:optimization}, following simulation results provided in Section \ref{sec:results}. Finally, Section \ref{sec:conclusion} concludes the paper.
    
\section{System Model}
\label{sec:system}
\begin{figure*} [t]
    \centering

\begin{tikzpicture}[
	sum/.style={circle, draw, fill=white, inner sep=0pt, minimum size=4mm},
	block/.style={rectangle, draw, dashed, minimum width=2.3cm, minimum height=2.3cm, fill=gray!5, inner sep=0pt, rounded corners},
	pa/.style = {regular polygon, regular polygon sides=3,
		draw, fill=white,shape border rotate=-90, minimum size=13mm},
	fiber/.style={rectangle, draw, minimum width=1cm, minimum height=0.5cm, inner sep=0pt,, fill=white},
	func/.style={rectangle, draw, minimum width=0.8cm, minimum height=0.5cm, fill=white, inner sep=0pt},
	line/.style={draw, -latex},
	]
	
	\newcommand{\multiplier}[3]{
		\begin{scope}[shift={(#2)}]
			\node[circle, inner sep=0pt, minimum size=4mm, fill=white] (#1) at (0,0) {}; 
			\draw (0,0) circle (2mm); 
			\draw (-0.14, -0.14) -- (0.14, 0.14); 
			\draw (-0.14, 0.14) -- (0.14, -0.14); 
			\node at (0,0) {#3}; 
		\end{scope}
	}
	
	\node (input) at (0,0) {$x^{(0)}_n$};
	
	\newcommand{\sumposxz}{1}
        \node[sum] (sum0) at (\sumposxz,0) {+};
	\node (noise) at (\sumposxz,-1) {$\sigma w^{(1)}_n$};	
	\path[line](input) -- (sum0);
        \path[line](noise) -- (sum0);
	
        \newcommand{\blockposx}{\sumposxz+2.3}
	\newcommand{\funposx}{\blockposx-0.5}
	\newcommand{\pagainposx}{\blockposx+0.6}
	\newcommand{\blockposy}{-0.1}
        \newcommand{\blocksepx}{2.5}
	\node[block] (block) at (\blockposx,\blockposy) {};
	\node(panum) at (\blockposx, 0.7) {PA-1};
	\node[func](pa) at (\funposx, 0){$f(\cdot)$};
	\multiplier{pagain}{\pagainposx, 0}{};
	\node (pagainval) at (\pagainposx,-1) {$g^{(1)}$};
	\path[line](pagainval) --(pagain);
	\path[line](sum0) -- (pa)node[midway, above]{$x^{(1)}_n$};
	\path[line](pa) -- (pagain);
	
	\newcommand{\sumposx}{\blockposx+\blocksepx}
        \node[sum] (sum) at (\sumposx,0) {+};
	\node (noise) at (\sumposx,-1) {$\sigma w^{(2)}_n$};	
	\path[line](noise) -- (sum);
        \path[line](pagain) -- (sum)node[midway, above]{$y^{(1)}_n$};
	
        \newcommand{\blockposxx}{\sumposx+\blocksepx}
	\newcommand{\funposxx}{\blockposxx-0.5}
	\newcommand{\pagainposxx}{\blockposxx+0.6}
	\node[block] (block2) at (\blockposxx,\blockposy) {};
	\node(panum2) at (\blockposxx, 0.7) {PA-2};
	\node[func](pa2) at (\funposxx, 0){$f(\cdot)$};
	\multiplier{pagain2}{\pagainposxx, 0}{};
	\node (pagainval2) at (\pagainposxx,-1) {$g^{(2)}$};
	\path[line](pagainval2) --(pagain2);
	\path[line](pa2) -- (pagain2);
        \path[line](sum) -- (pa2)node[midway, above]{$x^{(2)}_n$};  

	\newcommand{\sumposxx}{\blockposxx+\blocksepx}
        \node[sum] (sum2) at (\sumposxx,0) {+};
	\node (noise2) at (\sumposxx,-1) {$\sigma w^{(3)}_n$};	
	\path[line](noise2) -- (sum2);
        \path[line](pagain2) -- (sum2)node[midway, above]{$y^{(2)}_n$};
	
	\draw[-] (sum2) -- (\sumposxx+0.8, 0);
	\draw [dotted] (\sumposxx+0.8,0) -- (\sumposxx+1.3,0);
	
        \newcommand{\blockposxxx}{\sumposxx+1.35*\blocksepx}
	\newcommand{\funposxxx}{\blockposxxx-0.5}
	\newcommand{\pagainposxxx}{\blockposxxx+0.6}
	\node[block] (block3) at (\blockposxxx,\blockposy) {};
	\node(panum3) at (\blockposxxx, 0.7) {PA-$K$};
	\node[func](pa3) at (\funposxxx, 0){$f(\cdot)$};
	\multiplier{pagain3}{\pagainposxxx, 0}{};
	\node (pagainval3) at (\pagainposxxx,-1) {$g^{(K)}$};
	\path[line](pagainval3) --(pagain3);
	\path[line](pa3) -- (pagain3);
        \path[line](\sumposxx+1.3,0) -- (pa3)node[midway, above]{$x^{(K)}_n$};  
	
	\node (output) at (\blockposxxx+2.2,0) {$y^{(K)}_n$};
	\path[line](pagain3) -- (output);

	


	
\end{tikzpicture}
    \caption{Model of the cascaded PAs.}
    \label{fig:cascaded_pas_model}
\end{figure*}
Due to inherent nonlinear behavior of PAs in practice, they introduce nonlinear distortion to the incoming signal with the increase of power. 
To analyse the effect of total nonlinearities in the cascaded structure, where the distortions are accumulated with the increase in the number of amplifiers, the cascaded PAs can be modeled as a cascade of nonlinear functions indicating the PA behavior and noise sources representing thermal noise, as it is shown in Fig. \ref{fig:cascaded_pas_model}. 

We assume that the complex baseband equivalent of the input signal to the first PA in the cascade is denoted by $x_n^{(0)}$, with power of $p_0$.
PA behavior under normal operation conditions, in the simplest case, can be modeled using the third-order memoryless polynomial model \cite{Moghaddam_2022}. Thus, the output signal $y_n^{(k)}$ of the $k$th PA can be described as 
\begin{align}
 y^{(k)}_n&=g^{(k)}f\big(x_n^{(k)}\big)\notag\\
 & = g^{(k)}\big(x_n^{(k)}+\alpha^{(k)} x_n^{(k)}|x_n^{(k)}|^2\big),
 \label{eq:PA_model}
\end{align}
where $x_n^{(k)}$ is the input signal to the $k$th PA, $\alpha^{(k)}$ is the third-order nonlinearity coefficient, which is typically complex-valued to capture both amplitude-to-amplitude modulation (AM/AM) and amplitude-to-phase modulation (AM/PM). 
Since third-order nonlinearities are generally dominant compared to higher-order terms in the context of a single PA \cite{Moghaddam_2022}, the model in \eqref{eq:PA_model} captures the majority of nonlinear distortions and can be used for further analysis of a cascade of PAs. 
Additionally, in this paper, we assume that the amplifiers are placed in cascade to compensate for the losses incurred in the connectors between them. Therefore, the coefficient $g^{(k)}$ represents both linear amplification of the PA and attenuation due to losses in the connector. In the case when linear amplification exactly equals attenuation, $g^{(k)}=1$.  
Furthermore, the thermal noise between the PAs is modeled as additive white Gaussian noise (AWGN) with the variance $\sigma^2$.

To express the output of the $(k+1)$th PA, a function incorporating a single stage, represented by the PA nonlinearity and thermal noise, is defined as
\begin{align}
    F(y_n^{(k)}, g^{(k+1)}) = g^{(k+1)}f\big(y_n^{(k)} +\sigma w_n^{(k+1)}\big).
    \label{eq:funF}
\end{align}

Then, for the signal $x_n^{(0)}$ passing through $K$ cascaded PAs, the output $y^{(K)}_n$ can be expressed by applying the function $F(\cdot)$ in \eqref{eq:funF} iteratively $K$ times, i.e.,
\begin{align}
    y^{(K)}_n
    = \underbrace{F(F(\dots F(x_n^{(0)}, g^{(1)}), g^{(2)})\dots)}_{K \text{ times}}.
    \label{eq:Funcion_K}
\end{align}

\section{Analysis of Nonlinearities in the Cascaded Structure}
\label{sec:analysis}
In this section, we focus on the analysis of nonlinearities accumulating in the cascaded structure to show potential ways for minimizing the total distortions by optimizing parameters of the cascaded structure, as will be presented in Section \ref{sec:optimization}.
\subsection{Cascaded Nonlinearities in the Noise-Free Case}
\label{sec:analysis_noise_free}
In the noise-free case ($\sigma=0$), 
the output of the second PA can be written in terms of $x_n^{(0)}$ using \eqref{eq:Funcion_K} as
\begin{align}
    y_n^{(2)} =& g^{(2)} \bigg[g^{(1)} \big(x_n^{(0)}+\alpha^{(1)} x_n^{(0)}|x_n^{(0)}|^2\big) \notag \\
&+ \alpha^{(2)}\big(g^{(1)}\big)^3 \big(x_n^{(0)}+\alpha^{(1)} x_n^{(0)}|x_n^{(0)}|^2\big)\notag\\
    &\times \big|x_n^{(0)}+\alpha^{(1)} x_n^{(0)}|x_n^{(0)}|^2\big|^2\bigg].
    \label{eq:PA_out_k1}
\end{align}

Assuming that $\alpha^{(k)}$, $k=1,\dots, K$, are small values, we can approximate the quadratic terms $\alpha^{(k)}\alpha^{(l)}$ for $k, l=1,2, \dots, K$ as being negligibly small. Consequently, the terms in \eqref{eq:PA_out_k1} containing these coefficients can be omitted from this analysis. Then, $y_n^{(2)}$ in \eqref{eq:PA_out_k1} can be approximated by $\tilde{y}_n^{(2)}$ given by
\begin{align}
    \tilde{y}_n^{(2)} &= g^{(2)}g^{(1)}
    \bigg[ x_n^{(0)}+\bigg(\alpha^{(1)}+\alpha^{(2)}\big(g^{(1)}\big)^2\bigg)x_n^{(0)}|x_n^{(0)}|^2\bigg].
    \label{eq:PA_out_k1_simpl}
\end{align}

Applying the same calculations and utilizing the assumption $\alpha^{(k)}\alpha^{(l)}\approx0$ for $k, l=1,2, \dots, K$, the output of the third PA can be expressed as
\begin{align}
    \tilde{y}_n^{(3)} = & g^{(3)}g^{(2)}g^{(1)}
    \bigg[ x_n^{(0)}+\bigg(\alpha^{(1)}+\alpha^{(2)}\big(g^{(1)}\big)^2\notag\\
    & +\alpha^{(3)}\big(g^{(2)}g^{(1)}\big)^2\bigg)x_n^{(0)}|x_n^{(0)}|^2\bigg].
    \label{eq:PA_out_k2_simpl}
\end{align}

Following the same procedure, the $K$th PA output can be approximated by
\begin{align}
    \tilde{y}_n^{(K)} =\tilde{g}(\mathbf{g}, K) \Big[x_n^{(0)}+\tilde{\alpha}(\mathbf{g}, \boldsymbol{\alpha}, K)x_n^{(0)}|x_n^{(0)}|^2\Big],
    \label{eq:PA_K}
\end{align}
where $\tilde{g}(\mathbf{g}, K)$ and $\tilde{\alpha}(\mathbf{g}, \boldsymbol{\alpha}, K)$ can be seen as the gain and third-order nonlinearity coefficient of an equivalent PA, given respectively by
\begin{align}
    \tilde{g}(\mathbf{g}, K) &= \prod_{k=1}^{K} g^{(k)}, \label{eq:gain_approx_K}\\
    \tilde{\alpha}(\mathbf{g}, \boldsymbol{\alpha}, K) & =\alpha^{(1)}+ \sum_{k=2}^{K}\alpha^{(k)}\prod_{q=1}^{k-1}\big(g^{(q)}\big)^2, \label{eq:alpha_approx_K}
\end{align}
with $\mathbf{g}=[g^{(1)}, g^{(2)}, \dots, g^{(K)}]$ and $\boldsymbol{\alpha}=[\alpha^{(1)}, \alpha^{(2)}, \dots, \alpha^{(K)}]$.

\subsection{Cascaded Nonlinearities in the Presence of Noise}
\label{sec:analysis_noise}
In the presence of noise with variance $\sigma^2$, the input to the $k$th PA, $k=1,2, \dots, K$, contains the additional component $\sigma w_n^{(k)}$. Then, following the same analysis as in Section  \ref{sec:analysis_noise_free} and assuming that $\sigma$ is also a small value, the quadratic terms $\alpha^{(k)}\alpha^{(l)}$, $\alpha^{(k)}\sigma$, $\sigma^2$, for $k, l=1,2, \dots, K$, can be approximated as being negligibly small. Then, the output of the $K$th PA can be approximated by
\begin{align}
    \tilde{y}_n^{(K)}=&\tilde{g}(\mathbf{g}, K) \Big[x_n^{(0)}+\tilde{\alpha}(\mathbf{g}, \boldsymbol{\alpha}, K)x_n^{(0)}|x_n^{(0)}|^2\Big] \notag\\
    &+ \tilde{\sigma}(\mathbf{g}, K)w_n,
    \label{eq:PA_K_noise}
\end{align}
where $\tilde{\sigma}(\mathbf{g}, \mathbf{a}, K)$ is the standard deviation of an equivalent AWGN noise after the $K$th PA given by
\begin{align}
     \tilde{\sigma}(\mathbf{g}, K)=\sigma\sqrt{\sum_{k=1}^{K}\prod_{q=k}^{K}(g^{(q)})^2}.
     \label{eq:sigma_approx_K}
\end{align}

\subsection{Analysis of the Approximating Expressions}
\label{sec:analysis_discussions}
From equations \eqref{eq:PA_K}--\eqref{eq:sigma_approx_K}, it can be observed that the amount of distortions at the output of the cascade of $K$ PAs strongly depend on their parameters $g^{(k)}$ and $\alpha^{(k)}$, $k=1, \dots, K$, and \new{the values of the input signal $x_n^{(0)}$ to the first stage, i.e., the power $p_0$.}
Specifically, a linear combination of $\alpha^{(k)}$ in \eqref{eq:alpha_approx_K} relies on the product of quadratic terms $(g^{(k)})^2$, and noise level in \eqref{eq:sigma_approx_K} is influenced by the product of the linear terms $g^{(k)}$.
The coefficient $\alpha^{(k)}$ represents the inherent nonlinear distortion characteristics of the $k$th amplifier, that is generally difficult to change without redesigning the amplifier. In contrast, the input power $p_0$ and linear gain $g^{(k)}$ can be more easily adjusted in practice.
In this paper, we consider that the PAs in cascade have the same nonlinear characteristics, i.e.,  $\alpha^{(k)}=\alpha$. Thus, varying the parameters $p_0$ and $g^{(k)}$ can lead to minimizing the total nonlinearities at the output of the $K$th PA.

One of the challenges lies in finding a compromise between maintaining the desired linear gain at the output of the $K$th PA and minimizing the overall distortions because reducing the gain $g^{(k)}$ in \eqref{eq:alpha_approx_K} can also decrease the total linear amplification term in \eqref{eq:gain_approx_K} and can affect the noise component in \eqref{eq:sigma_approx_K}.

It is important to emphasize that equations \eqref{eq:PA_K}--\eqref{eq:sigma_approx_K} were derived to provide insights into the total nonlinearities and to highlight the need for optimizing the parameters of the cascaded structure in order to minimize the overall distortions. Further, the cascaded function in \eqref{eq:Funcion_K} will be utilized in the optimization problem presented in the next section.

\section{Optimization Problem}
\label{sec:optimization}
\subsection{Problem Formulation}
To tackle the problem of minimizing the total nonlinearities in the cascaded PAs structure, an NLS algorithm is employed in this paper. 

The problem is to minimize the total nonlinearities by optimizing the PA gains $g^{(k)}$ and input power $p_0$, i.e., 
\begin{align}
    &\min_{p_0, \mathbf{g}} &&  \lVert G x_n^{(0)}/\sqrt{p_0}-y_n^{(K)} \rVert ^2 \label{eq:optimization_problem_g}\\
    &\text{subject to } && 
    0 < p_0 \leq 1, \notag\\
    & && (1-\varepsilon)G \leq g^{(k)} \leq (1+\varepsilon)G, k=1,2, \dots, K,
    \label{eq:optimization_problem_g_constraints}
\end{align}
where $G$ is the reference gain (desired) at the output of the $K$th PA, $y_n^{(K)}$ as in \eqref{eq:Funcion_K}, 
and $\varepsilon$ is the parameter determining the allowable range for the PA gains $g^{(k)}$. 
Here, we consider two following sub-cases, that add additional constraints to \eqref{eq:optimization_problem_g_constraints}:
\begin{enumerate}
    \item[(a)] Equal PA gains, i.e., $g^{(k)}=g$ and $\mathbf{g}=[g, g, \dots, g]$. Then, the optimization parameters are $p_0$ and $g$ in \eqref{eq:optimization_problem_g}.
    \item[(b)] Unequal PA gains, i.e., the optimization parameters are $p_0$ and $\mathbf{g}=[g^{(1)}, g^{(2)}, \dots, g^{(K)}]$ in \eqref{eq:optimization_problem_g}.
\end{enumerate}
Note that the normalization by $1/\sqrt{p_0}$ in \eqref{eq:optimization_problem_g} eliminates the dependence of the desired signal on the input power $p_0$.

\subsection{Performance Metrics}
The performance of the cascaded PA models with optimized parameters is evaluated in terms of normalized mean square error (NMSE) and adjacent channel leakage ratio (ACLR). 
The NMSE is defined as
\begin{align}
    \text{NMSE} = \frac{\sum_{n=0}^{N}|y_{n}-y_n^{(K)}|^2}{\sum_{n=0}^{N}|y_{n}|^2},
    \label{eq:nmse}
\end{align}
where $y_{n}=Gx_n^{(0)}/\sqrt{p_0}$ is the desired output and $y_n^{(K)}$ is the output of the $K$th PA in \eqref{eq:Funcion_K}.

The ACLR is defined as
\begin{align}
    \text{ACLR} = \frac{\int_{\text{max. adj.}}\big|Y^{(K)}(f)\big|^2}{\int_{\text{ch.}}\big|Y^{(K)}(f)\big|^2},
    \label{eq:aclr}
\end{align}
where $Y^{(K)}(f)$ is the power spectrum of the signal $y_n^{(K)}$. Here, the integration in the nominator is performed over the adjacent channel with the highest power, while the denominator's integration is carried out over the main channel.

\section{Simulation Results}
\label{sec:results}
\subsection{The Choice of the Starting Point for the Optimization}
\begin{figure} [t]
    \begin{minipage}{0.5\textwidth}
        \centering
        \includegraphics[trim={1.6cm 0.2cm 2cm 0.2cm},clip,width=\linewidth]{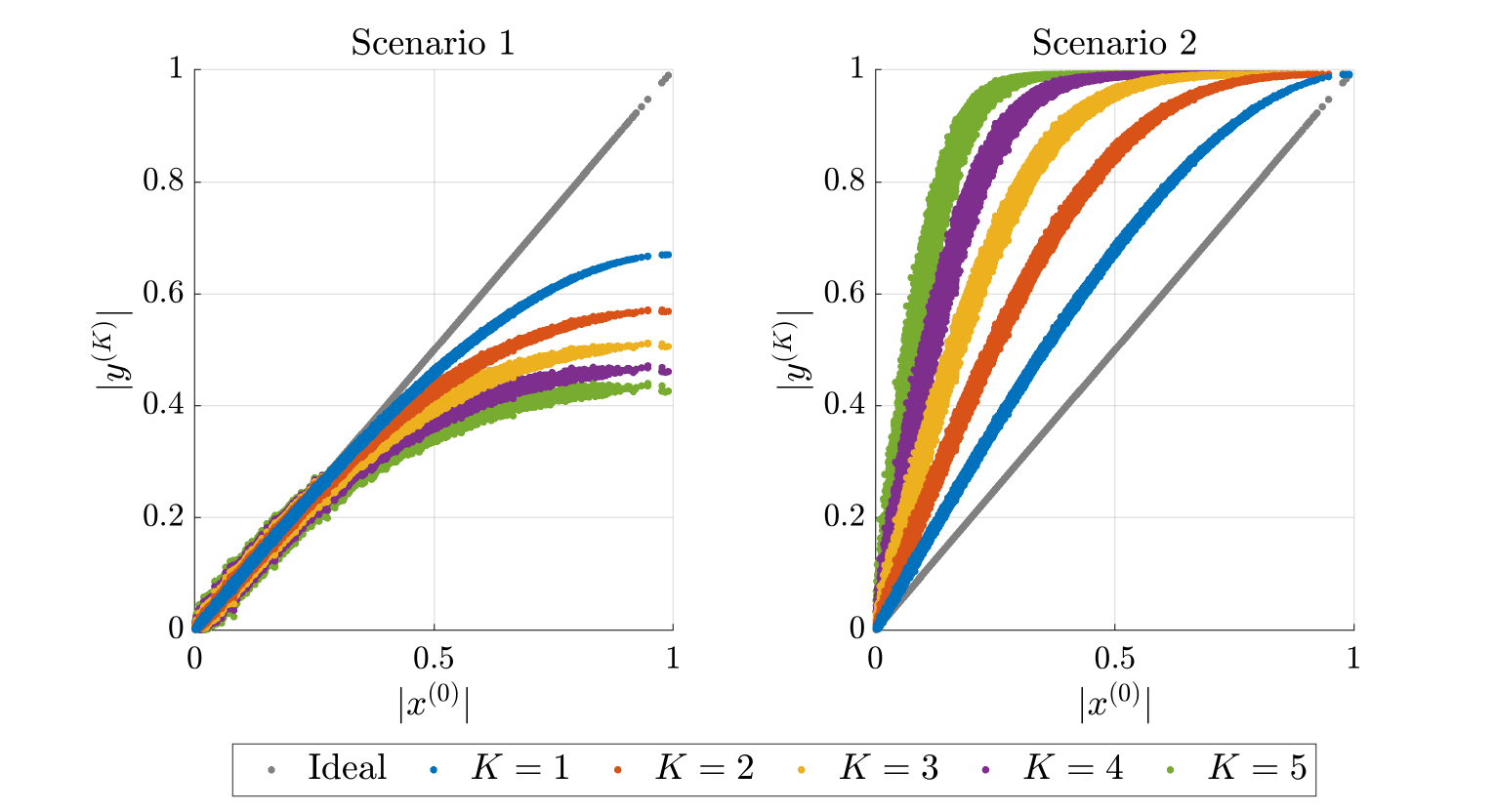}
        \caption{\new{AM/AM of the cascaded PAs for the two initial scenarios.}}
        \label{fig:InitialScenarios-AMAM}
        \vspace{0.1cm}
    \end{minipage}
    \begin{minipage}{0.5\textwidth}
        \centering
        \includegraphics[trim={1.6cm 0.2cm 2cm 0.2cm},clip,width=\linewidth]{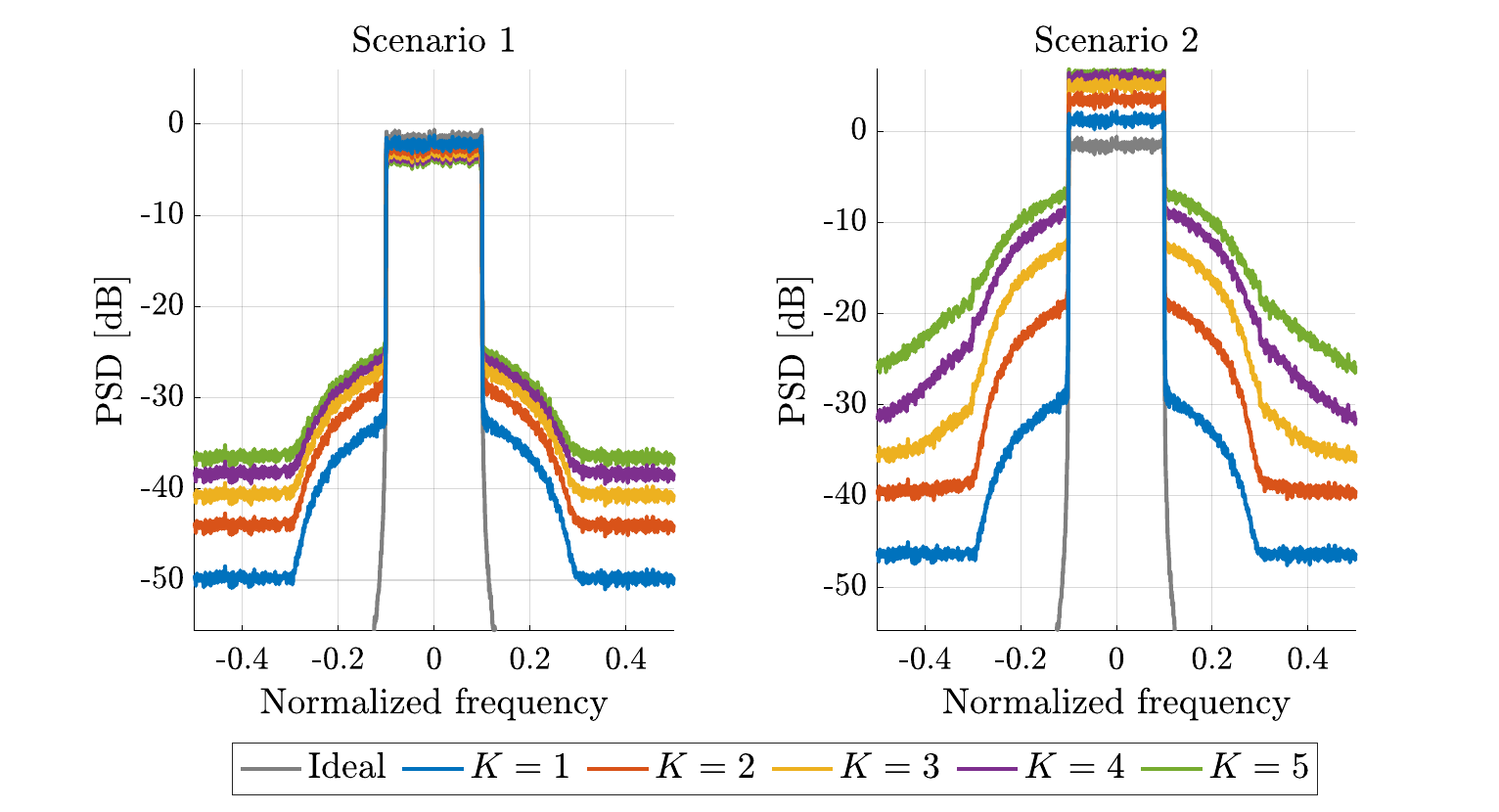}
        \caption{\new{PSD of the cascaded PAs for the two initial scenarios.}}
        \label{fig:InitialScenarios-spec}
    \end{minipage}
\end{figure}
Since the optimization problem in \eqref{eq:optimization_problem_g} is nonlinear, a reasonable starting point is needed to prevent the convergence to a poor local optimum. Thus, we consider two initial scenarios:
\begin{enumerate}
    \item[(1)] The linear gain of every PA corresponds to the losses that occurred in the connector, i.e., $g^{(k)}=1$, $k=1, 2,\dots, K$.
    \item[(2)] Every PA operates in the most efficient regime, i.e., with the maximum input power, and in this case, $g^{(k)}=g=x_{\text{max}}/f(x_{\text{max}})$ representing amplification to achieve the maximum input level of one, with $x_{\text{max}}=\sqrt{1/(3|\alpha|)}$.
\end{enumerate}
A cascade of $K=1, \dots, 5$ PA(s) is considered for both scenarios, with 
\new{$\alpha=-0.33(1-0.1j)$}, $p_0 = 1$ and \new{$\sigma^2=10^{-5}$.}
The input signal $x_n^{(0)}$ is a normalized $16\text{QAM}$ complex baseband signal. The desired linear gain $G$ is set to one. The AM/AM characteristics and the power spectral density (PSD) of the $K$th PA output signals for the two initial scenarios are shown in Figs. \ref{fig:InitialScenarios-AMAM} and \ref{fig:InitialScenarios-spec}, respectively. Clearly, as 
$K$ increases, the level of distortion rises. In Scenario 1, the output signal of each successive PA accumulates nonlinear distortions and experiences saturation, resulting in a progressive loss of the initial amplitude as the number of cascaded PAs increases. In Scenario 2, the output signal becomes increasingly susceptible to nonlinear distortions due to enhanced nonlinear effects, while its amplitude remains approximately unchanged. 
The results in terms of NMSE and ACLR 
are shown in Fig. \ref{fig:ResultsMetrics}, where it seen that the first scenario shows better performance, and thus can serve as the initial starting point for further optimization. Moreover, these two scenarios can serve as the upper and lower bounds for choosing a starting point to solve the nonlinear optimization problem in \eqref{eq:optimization_problem_g}. 

\begin{table}[t]
\begin{minipage}{0.5\textwidth}
\centering
\renewcommand{\arraystretch}{1.1}
\caption{\new{Input Signal Power After Optimization.}}
\label{tab:power_opt}
\begin{tabular}{p{3.5cm} | c |c|c|c|c}
    \hline
    & \multicolumn{5}{c}{$K$ PAs in a cascade}\\
    Case & $1$ &  $2$ & $3$ & $4$ & $5$\\
    \hline
    Input power opt. (Scenario 1) & 1 & 1 & 1 & 1 & 1\\
    \hline 
    Input power opt. (Scenario 2) 
    & 0.49 & 0.23 & 0.11 & 0.05 & 0.02\\
    \hline
    Input power and equal PA gains optimization 
    & 0.65 & 0.40 & 0.25 & 0.19 & 0.19\\
    \hline 
    Input power and unequal PA gains optimization 
    & 0.65 & 0.40 & 0.27 & 0.30 & 0.32\\
    \hline
\end{tabular}
\end{minipage}
\begin{minipage}{0.5\textwidth}
\centering
\renewcommand{\arraystretch}{1.1}
\caption{\new{PA Gain Values After Optimization.}}
\label{tab:gains_opt}
\begin{tabular}{p{3.3cm} | c|c|c|c|c|c}
    \hline
    & & \multicolumn{5}{c}{$K$ PAs in a cascade}\\
    Case & $k$ &$1$ &  $2$ & $3$ & $4$ & $5$\\
    \hline
    \multirow{5}{=}{Equal PA gains optimization} 
    & $1$ & 1.07  & 1.07  & 1.07  & 1.06  & 1.06  \\  
    & $2$ & -     & 1.07  & 1.07  & 1.06  & 1.06  \\  
    & $3$ & -     & -     & 1.07  & 1.06  & 1.06  \\  
    & $4$ & -     & -     & -     & 1.06  & 1.06  \\  
    & $5$ & -     & -     & -     & -     & 1.06  \\  
    \hline
    \multirow{5}{=}{Unequal PA gains optimization} 
    & $1$ & 1.07  & 0.86  & 0.70  & 0.70  & 0.70  \\  
    & $2$ & -     & 1.30  & 1.26  & 0.99  & 0.82  \\  
    & $3$ & -     & -     & 1.30  & 1.30  & 1.21  \\  
    & $4$ & -     & -     & -     & 1.30  & 1.30  \\  
    & $5$ & -     & -     & -     & -     & 1.30  \\ 
    \hline
    \multirow{5}{=}{Input power and equal PA gains optimization} 
    & $1$ & 1.30  & 1.30  & 1.30  & 1.26  & 1.21  \\  
    & $2$ & -     & 1.30  & 1.30  & 1.26  & 1.21  \\  
    & $3$ & -     & -     & 1.30  & 1.26  & 1.21  \\  
    & $4$ & -     & -     & -     & 1.26  & 1.21  \\  
    & $5$ & -     & -     & -     & -     & 1.21  \\   
    \hline
    \multirow{5}{=}{Input power and unequal PA gains optimization} 
    & $1$ & 1.30  & 1.30  & 1.24  & 1.02  & 1.02  \\  
    & $2$ & -     & 1.30  & 1.30  & 1.17  & 1.03  \\  
    & $3$ & -     & -     & 1.30  & 1.30  & 1.13  \\  
    & $4$ & -     & -     & -     & 1.30  & 1.29  \\  
    & $5$ & -     & -     & -     & -     & 1.30  \\  
    \hline
\end{tabular}
\vspace{-0.1cm}
\end{minipage}
\end{table}

\subsection{Optimization of the Parameters in the Cascaded Structures}
\subsubsection{Input Power Optimization}
First, to investigate how the input power $p_0$ influences the performance, the problem in \eqref{eq:optimization_problem_g} is considered by optimizing a single coefficient $p_0$, 
given fixed PA gain coefficients $\textbf{g}$. The values of the optimized input signal power are listed in Table \ref{tab:power_opt}, and the performance of the optimized configurations are shown in Fig. \ref{fig:ResultsMetrics}. It is seen that $p_0=1$ is the optimal value for Scenario 1, while it is not for Scenario 2 where the performance has improved after optimization, and even outperforms Scenario 1. 
However, as the number of PAs in the cascade increases, the input power in Scenario 2 decreases significantly, potentially leading to a reduced SNR in the initial stages due to the constant thermal noise power.

\subsubsection{PA Gains Optimization}
Further, to investigate how the PA gains can be optimized given maximum input power $p_0=1$, the problem in \eqref{eq:optimization_problem_g} is considered with equal gains, i.e., by optimizing a single coefficient $g$, and with unequal gains, i.e., by optimizing coefficients $g^{(1)}, g^{(2)}, \dots, g^{(K)}$. Scenario 1 is chosen as the initial starting point, and the parameter $\varepsilon=0.3$ is set here.
The results shown in Fig. \ref{fig:ResultsMetrics} indicate that the case with unequal gains exhibits better performance. Notably, the difference between the two cases becomes more pronounced as the number of PAs in the cascade increases. The values of the optimized PA gain values are listed in Table \ref{tab:gains_opt}.

\subsubsection{Joint Input Power and PA Gains Optimization}
\begin{figure} [t]
\begin{minipage}{0.47\textwidth}
\centering
    \includegraphics[trim={1.6cm 0.2cm 2cm 0.2cm},clip,width=\linewidth]{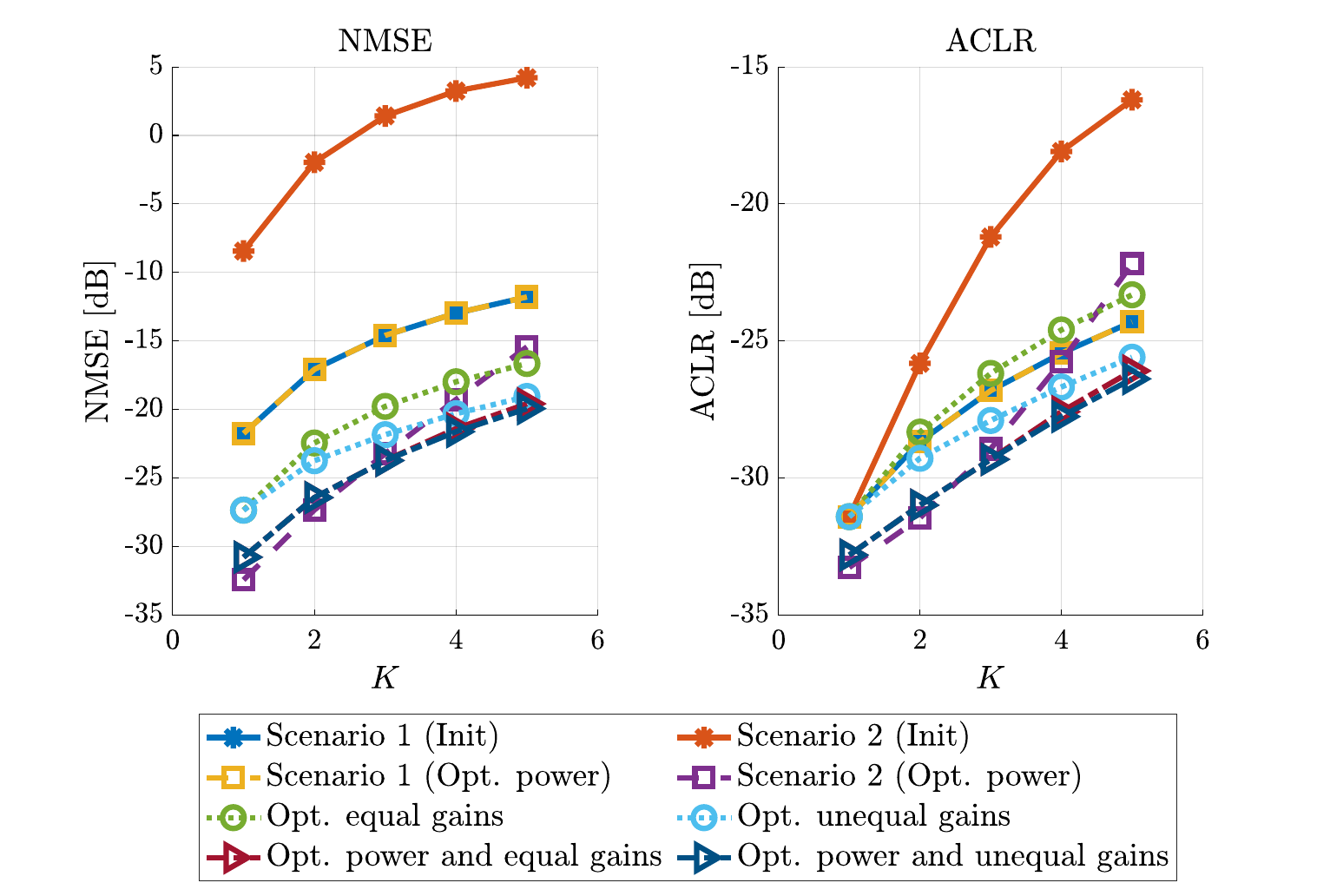}
    \caption{\new{Results of the optimization of the cascaded PAs parameters in terms of NMSE and ACLR.}}
    \label{fig:ResultsMetrics}
    \vspace{-0.15cm}
\end{minipage}
\begin{minipage}{0.47\textwidth}
        \vspace{0.2cm}
        \centering
        \includegraphics[trim={1.6cm 0.2cm 2cm 0.2cm},clip,width=\linewidth]{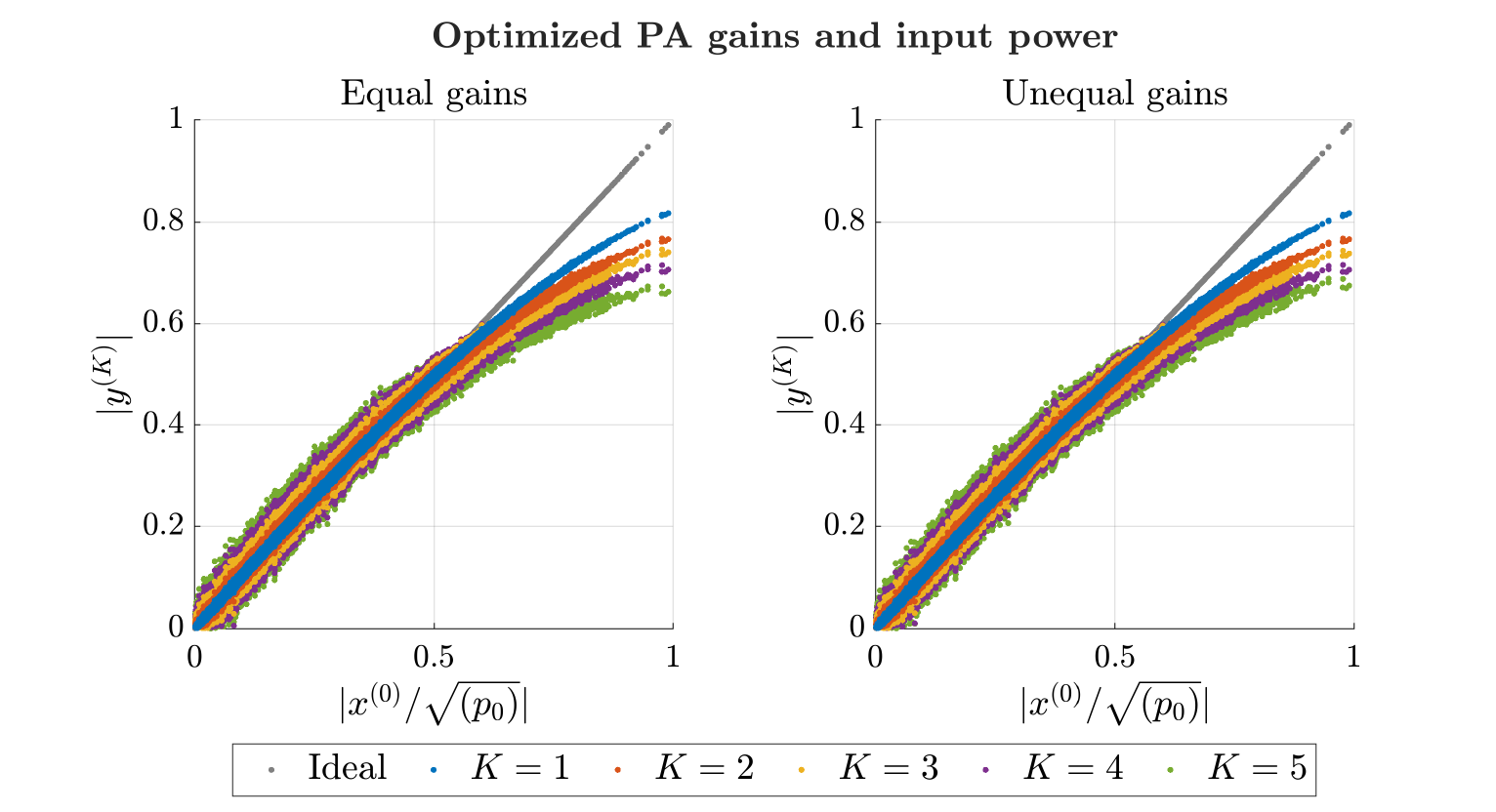}
        \caption{\new{AM/AM of the output after $K$ PAs with the optimized input power and PA gains (equal and unequal gains).}}
        \label{fig:ResOptGains-AMAM}
        \vspace{-0.15cm}
\end{minipage}
\begin{minipage}{0.47\textwidth}
        \vspace{0.2cm}
        \centering
        \includegraphics[trim={1.6cm 0.2cm 2cm 0.2cm},clip,width=\linewidth]{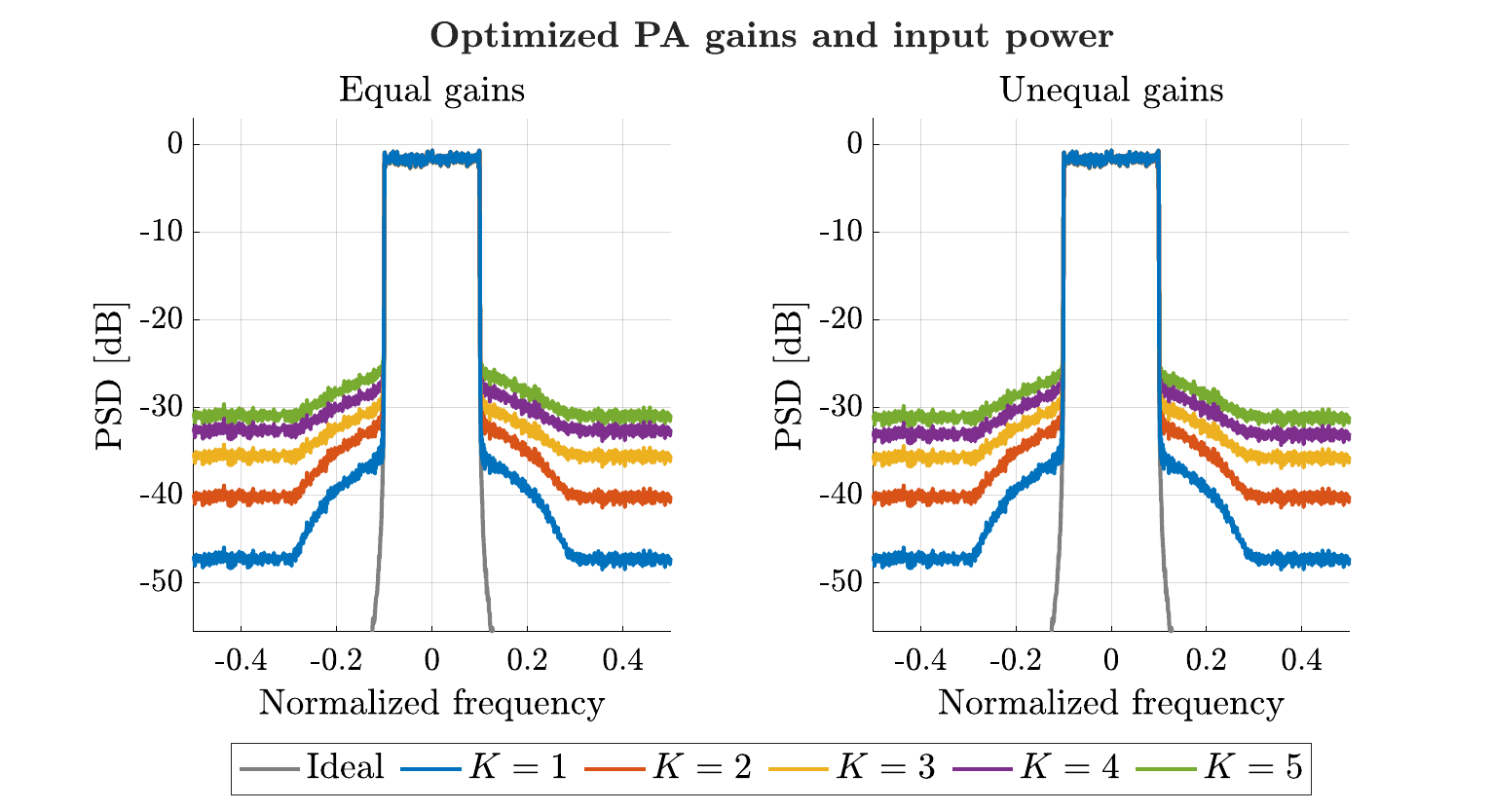}
        \caption{\new{PSD of the output after $K$ PAs with the optimized input power and PA gains (equal and unequal gains).}}
        \label{fig:ResOptGains-spec}
        \vspace{-0.2cm}
\end{minipage}
\end{figure}

Finally, joint optimization of the PA gains and input power is considered here. Scenario 1 as the starting point and $\varepsilon=0.3$ are also chosen here. The results of the optimization are shown in Figs. \ref{fig:ResultsMetrics}--\ref{fig:ResOptGains-spec}, and the optimized values of the input signal \new{power} and PA gains are listed in Tables \ref{tab:power_opt} and \ref{tab:gains_opt}, respectively.
One can observe that incorporating the input power level into the optimization, along with the gains, further enhances the potential for reducing overall distortions and mitigating the severe effects of cascading $K$ PAs, 
however, the difference between equal and unequal gains is small and becomes noticeable only for $K\geq4$.

\section{Conclusions}
\label{sec:conclusion}
This paper \new{studied modeling} 
of cascaded PAs and provided an analysis of the total nonlinearities, which accumulate as the number of PAs increases. The NLS algorithm was proposed to optimize PA gains and input power levels, thereby minimizing the overall impact of cascaded nonlinearities. Simulation results demonstrated the performance improvement of cascaded structures optimized using the proposed framework.

The analysis presented in this paper can provide opportunities to develop efficient linearization techniques while achieving an optimal balance between minimizing overall distortions, enhancing system performance, and reducing implementation complexity. In future work, higher-order polynomials can be utilized to model cascaded PAs more accurately, and digital predistortion techniques can be investigated to enhance the performance of cascaded PA structures.

\section{Acknowledgment}
\label{sec:acknowledgments}
This work was supported by the 6GTandem project funded by the European Union’s Horizon Europe research and innovation programme under Grant Agreement No 101096302.

\bibliographystyle{IEEEtran}
\bibliography{bibliography}

\end{document}